\documentclass[a4paper,twocolumn,prl,showpacs]{revtex4}
\usepackage{graphicx,amsmath,amssymb}
\renewcommand{\[}{\begin{equation}}
\renewcommand{\]}{\end{equation}}
\def\bea{\begin{eqnarray}}
\def\eea{\end{eqnarray}}
\def\nn{\nonumber\\}

\newcommand{\intr}{\int d{\bf r} \;}

\newcommand{\ei}[1]{{\rm e}^{i #1}}
\newcommand{\B}{{\bf B}}

\newcommand{\M}{{\bf M}}
\newcommand{\MM}{{\mathfrak M}}
\newcommand{\Mb}{{\bf M}_{\rm bulk}}
\newcommand{\Mc}{{\bf M}_{\rm cell}}
\newcommand{\m}{{\bf m}}
\newcommand{\CQ}{{\cal Q}}
\newcommand{\CP}{{\cal P}}

\newcommand{\ac}{A_{\rm cell}}
\newcommand{\ab}{A_{\rm bulk}}

\newcommand{\smu}{\sum_{\epsilon_j \le \mu}}
\newcommand{\smup}{\sum_{\epsilon_j > \mu}}
\renewcommand{\r}{{\bf r}}

\renewcommand{\P}{{\bf P}}

\newcommand{\equ}[1]{Eq.~(\ref{#1})}
\newcommand{\eqs}[2]{Eqs.~(\ref{#1}) and (\ref{#2})}

\def\bra#1{\langle#1\vert}
\def\ket#1{\vert#1\rangle}
\def\ev#1{\langle#1\rangle}
\def\me#1#2#3{\langle#1| \, #2 \, |#3\rangle}
\def\runtime{(\the\time)\qquad\the\month/\the\day/\the\year}
\def\today
 {\count10=\year\advance\count10 by -2000 \number\day--\ifcase
  \month \or Jan\or Feb\or Mar\or Apr\or May\or Jun\or
             Jul\or Aug\or Sep\or Oct\or Nov\or Dec\fi--\number\count10}

\def\hour{\count10=\time\count11=\count10
\divide\count10 by 60 \count12=\count10
\multiply\count12 by 60 \advance\count11 by -\count12\count12=0
\number\count10 :\ifnum\count11 < 10 \number\count12\fi\number\count11}

\begin{document}

\title{Irrelevance of the boundary on the magnetization of metals}

\author{Antimo Marrazzo$^{1,2}$ and Raffaele Resta$^{2,3}$ \\ ~~}

\affiliation{\centerline{$^1$\footnote{Present address} Theory and Simulation of Materials (THEOS), \`Ecole Polytechnique F\'ed\'erale de Lausanne, CH-1015 Lausanne, Switzerland}}

\affiliation{\centerline{$^2$\footnote{Permanent address} Dipartimento di Fisica, Universit\`a di Trieste, Strada Costiera 11, 34151 Trieste, Italy}}

\affiliation{\centerline{$^3$ Donostia International Physics Center, 20018 San Sebasti\'an, Spain}}

\date{\today}

\begin{abstract}
The macroscopic current density responsible for the mean magnetization $\M$ of a uniformly magnetized bounded sample is localized near its surface. In order to evaluate $\M$ one needs the current distribution in the whole sample: bulk and boundary. In recent years it has been shown that the boundary has no effect on $\M$ {\it in insulators}: therein, $\M$ admits an alternative expression, not based on currents. $\M$ can be expressed in terms of the bulk electron distribution only, which is ``nearsighted'' (exponentially localized); this virtue is not shared by metals, having a qualitatively different electron distribution. We show, by means of simulations on paradigmatic model systems, that even in metals the $\M$ value can be retrieved in terms of the bulk electron distribution only.
\end{abstract}

\date{\today\ at \hour}


\maketitle \bigskip\bigskip

Electrical polarization $\P$ and orbital magnetization $\M$ share several properties; in the crystalline case $\P$ and $\M$ are both cast as Brillouin-zone integrals, where the integrands look similar. There is a key difference, though: $\P$ is defined modulo a ``quantum'' \cite{Vanderbilt93}, while $\M$ is not affected by such indeterminacy \cite{Xiao05,rap128,rap130,Souza08}. It follows that tinkering with the boundaries of a finite crystallite may affect $\P$, but not $\M$. How this could happen is far from obvious, since the circulating current responsible for $\M$ may be carried by edge states. It has been first shown by Chen and Lee in 2012 \cite{Chen12} that the total circulating current is indeed insensitive to boundary conditions; this finding was later exploited in Ref. \cite{rap148} in order to obtain an expression for $\M$ which is explicitly {\it local}; a similar result was independently found in Ref. \cite{Schulz13}. The literature so far \cite{Chen12,rap148,Schulz13,preprint} addresses insulators only, either trivial (Chern number $C=0$) or topological ($C \neq 0$). 

The extension to the metallic case is not obvious, since one of the reasons for the locality of $\M$ is the exponential decay of the one-body density matrix in insulators (``nearsightedness'' \cite{Kohn96}). In metals instead such decay is only power-law, which hints to a possibly different role of edge states. Furthermore, differently from insulators, in metals there are conducting states both in the bulk and at the surface.
In this Letter we give evidence that the above features do not spoil the locality of $\M$. Even in a metallic system $\M$ can be expressed in terms of the one-body density matrix in the bulk of the sample; tinkering with the boundaries cannot alter the $\M$ value. We show this by means of simulations on model two-dimensional (2D) Hamiltonians on finite samples within ``open'' boundary conditions (OBCs), where we break time-reversal symmetry in two different ways: either \`a la Haldane \cite{Haldane88}, or by means of a macroscopic $\B$ field.

We neglect any spin-dependent property here, dealing with ``spinless electrons''. The orbital dipole of a finite system of independent electrons at $T=0$ is \[ \m = - \frac{e}{2c} \smu \me{\varphi_j}{\r \times {\bf v}}{\varphi_j}, \label{def} \] where ${\bf v} = i [H,\r] /\hbar$ is the quantum-mechanical velocity operator, $\ket{\varphi_j}$ are the single-particle orbitals with energies $\epsilon_j$, and $\mu$ is the Fermi energy.
The macroscopic magnetization $\M$ is defined as the thermodynamic limit of $\m/V$, where $V$ is the system volume and the limit is taken at constant $\mu$. 
Ground state properties are expressed in terms of the density matrix (a.k.a. ground state projector) $\CP$; we will also need the complementary projector $\CQ= {\cal I} - \CP$. Their definitions are \[ \CP =  \smu \ket{\varphi_j} \bra{\varphi_j} , \quad \CQ = \smup \ket{\varphi_j} \bra{\varphi_j}  . \] 
The orbital dipole of the finite system, \equ{def}, is then \[  m_\gamma = - \frac{e}{2c} \epsilon_{\gamma\alpha\beta} \intr \ev{\r |\, r_\alpha v_\beta \, \CP \;| \r} ; \label{def1} \] according to Refs. \cite{Souza08,rap148,Schulz13,preprint} \equ{def1} is identically transformed into \[ m_\gamma = \frac{1}{2}  \epsilon_{\gamma\alpha\beta} \intr \MM_{\alpha\beta}(\r) , \label{exo} \] \bea \MM_{\alpha\beta}(\r) &=& \frac{e}{\hbar c} \mbox{Im }  \ev{\r |\, \CP r_\alpha \CQ (H-\mu) \CQ r_\beta \CP\, | \r} \nn &-& \frac{e}{\hbar c}  \mbox{Im }  \ev{\r |\, \CQ r_\alpha\CP(H-\mu) \CP  r_\beta \CQ \, | \r}  . \label{parts} \eea
In either insulating or metallic systems the integrated values provided by Eqs. (\ref{def}), (\ref{def1}), and (\ref{exo}) are identical, but the {\it integrands} therein are quite different. This is similar to what happens when integrating a function by parts; we also stress that any reference to microscopic currents has disappeared in \equ{parts}.

Only the insulating case has been addressed so far, where it has been proved \cite{Chen12,rap148,Schulz13,preprint} that \equ{exo} has the outstanding virtue of providing a {\it local} expression for $\M= \m/V$: instead of evaluating the trace over the whole system, as in \equ{exo}, we may evaluate the trace per unit volume in the bulk region of the sample. Notably, this converges (in the large system limit) {\it much faster} than the textbook definition based on \eqs{def}{def1}, where the boundary contribution to the integral is extensive (see also Fig. \ref{fig:ins} below). 

\begin{figure}[t]
\centering
\includegraphics[width=\columnwidth]{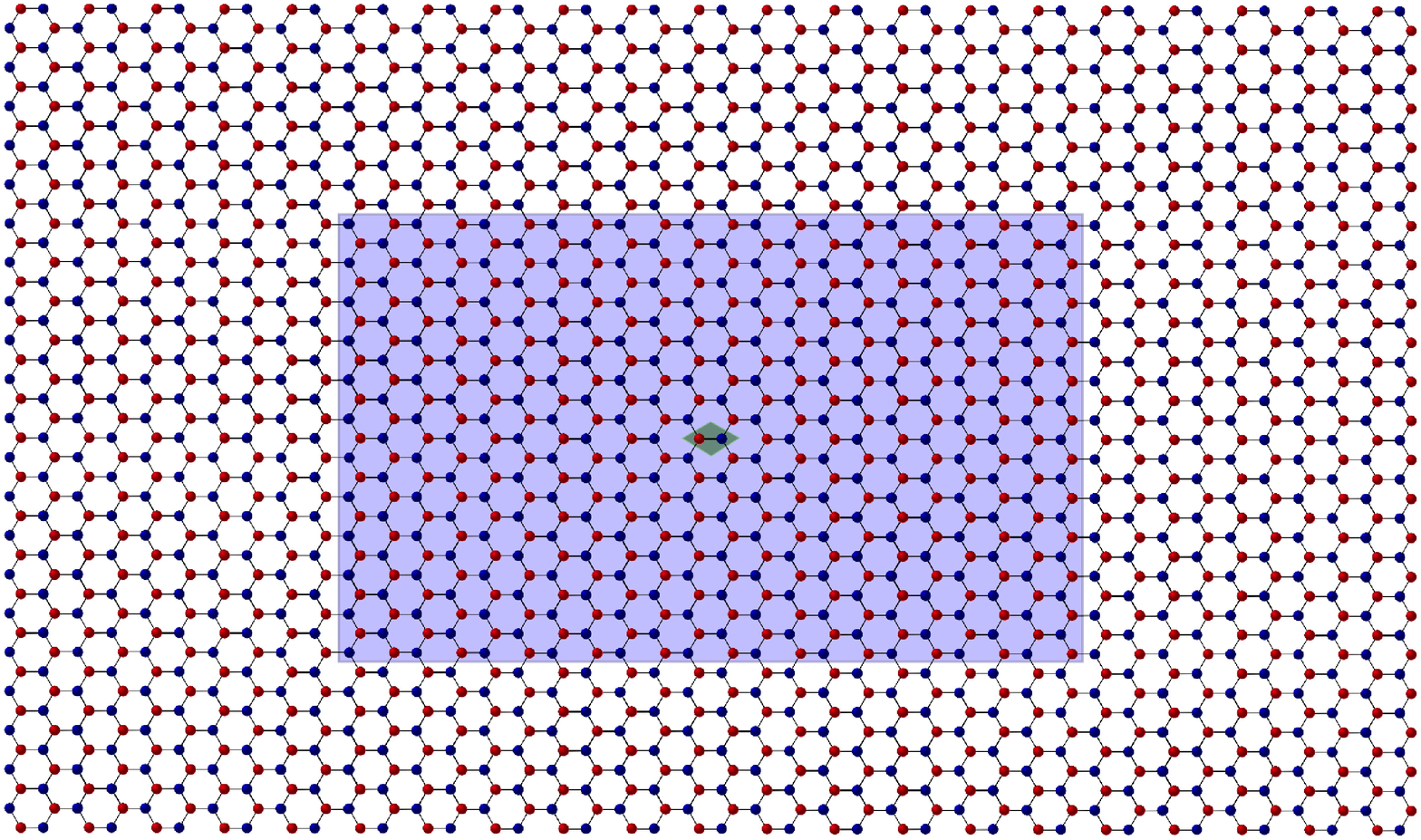}
\caption{(color online). A typical ``Haldanium'' flake. We have considered flakes with up to 8190 sites, all with the same aspect ratio; the one shown here has 1806 sites. In order to probe locality, the field $\MM(\r)$, \equ{parts}, is averaged either on the central cell (two sites) or on the ``bulk'' region (1/4 of the sites).}
\label{fig:flake} \end{figure}

The metallic case has not been addressed so far; in this work we investigate the behavior of $\MM(\r)$, \equ{exo}, in metallic 2D samples by means of simulations based on model tight-binding Hamiltonians. Our samples are finite flakes within OBCs, where the volume $V$ is replaced by area $A$. We remind that if instead one adopts periodic boundary conditions (PBCs), $\M$ has a known expression as a reciprocal-space integral \cite{rap130} which, however, only applies to magnetization in either vanishing or commensurate macroscopic $\B$ field. In this Letter we present OBCs test-case simulations for both $\B=0$ and $\B \neq 0$; the former case adopts rectangular flakes like the one shown in Fig. \ref{fig:flake}, while the latter adopts square flakes. For reasons thoroughly discussed below, the two cases present completely different features.

The paradigmatic model for breaking time-reversal symmetry without a macroscopic $\B$ field is the Haldane Hamiltonian \cite{Haldane88}, adopted here as well as by several authors in the past. Our choice of parameters is:  first- and second-neighbor hopping $t_1=1$ and $t_2 = \ei{\phi}/3$, with $\phi=0.25 \pi$; onsite energies $\pm \Delta $ with $\Delta = 1.5$.
With respect to the insulating case, the metallic one is computationally more demanding: in fact finite-size effects induce  large oscillations (as a function of the flake size) when the Fermi level $\mu$ is not in an energy gap. As usual, we deal with this problem by adopting the ``smearing'' technique: what we present here is the result of a combined large-size and small-smearing finite-size analysis. Here we adopt Fermi-Dirac smearing, although we stress that we are {\it not} addressing $\M$ at finite temperature \cite{Xiao06,Shi07}: the smearing is a mere computational tool.

\begin{figure}[t]
\centering
\includegraphics[width=0.8\columnwidth]{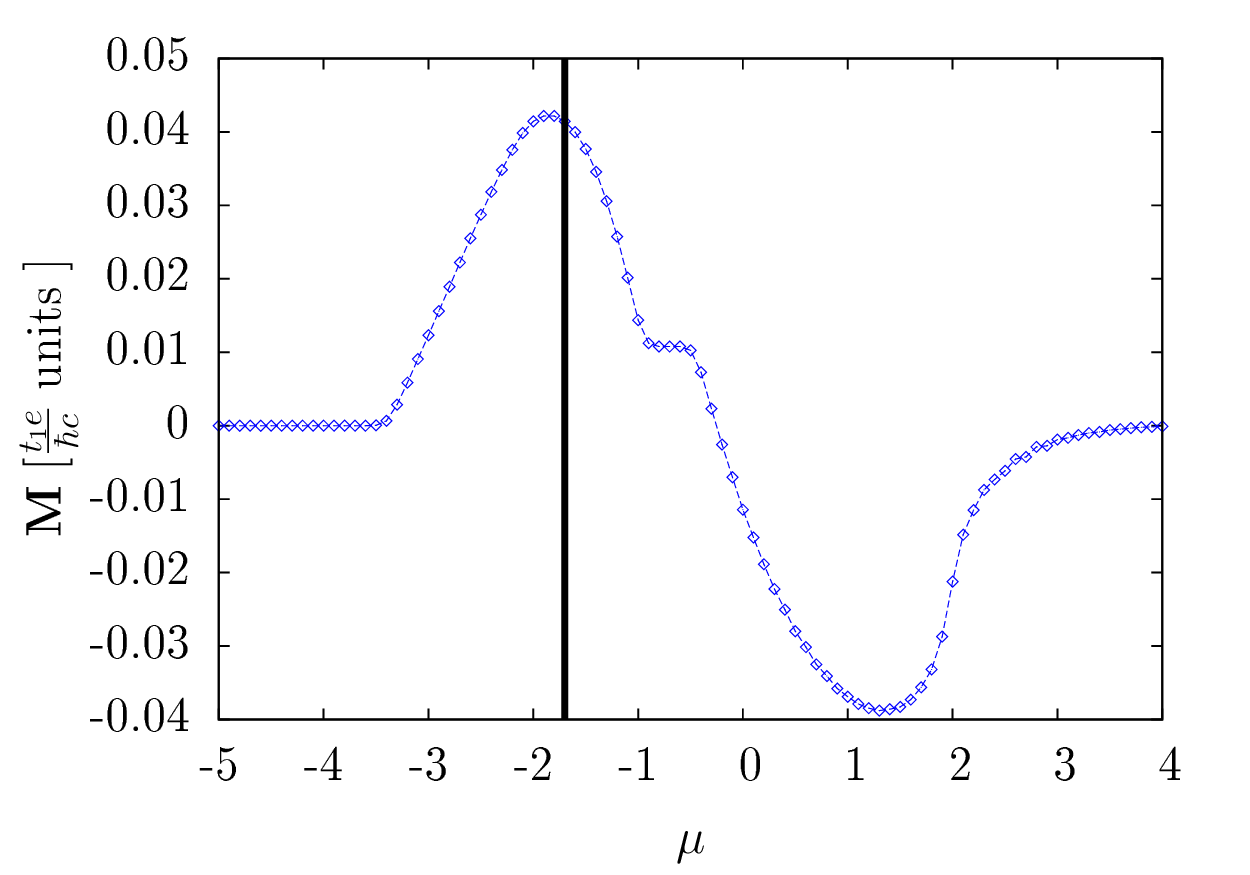} 
\caption{(color online). The magnetization of a large flake (6162 sites) as a function of the Fermi level $\mu$.
The valence-conduction gap is between $\varepsilon=-0.4$ and  $\varepsilon=-1.0$; our metallic simulations are at $\mu=-1.7$, shown as a vertical line.}
\label{fig:sweep1} \end{figure}

\begin{figure}[b]
\centering
\includegraphics[width=.8\columnwidth]{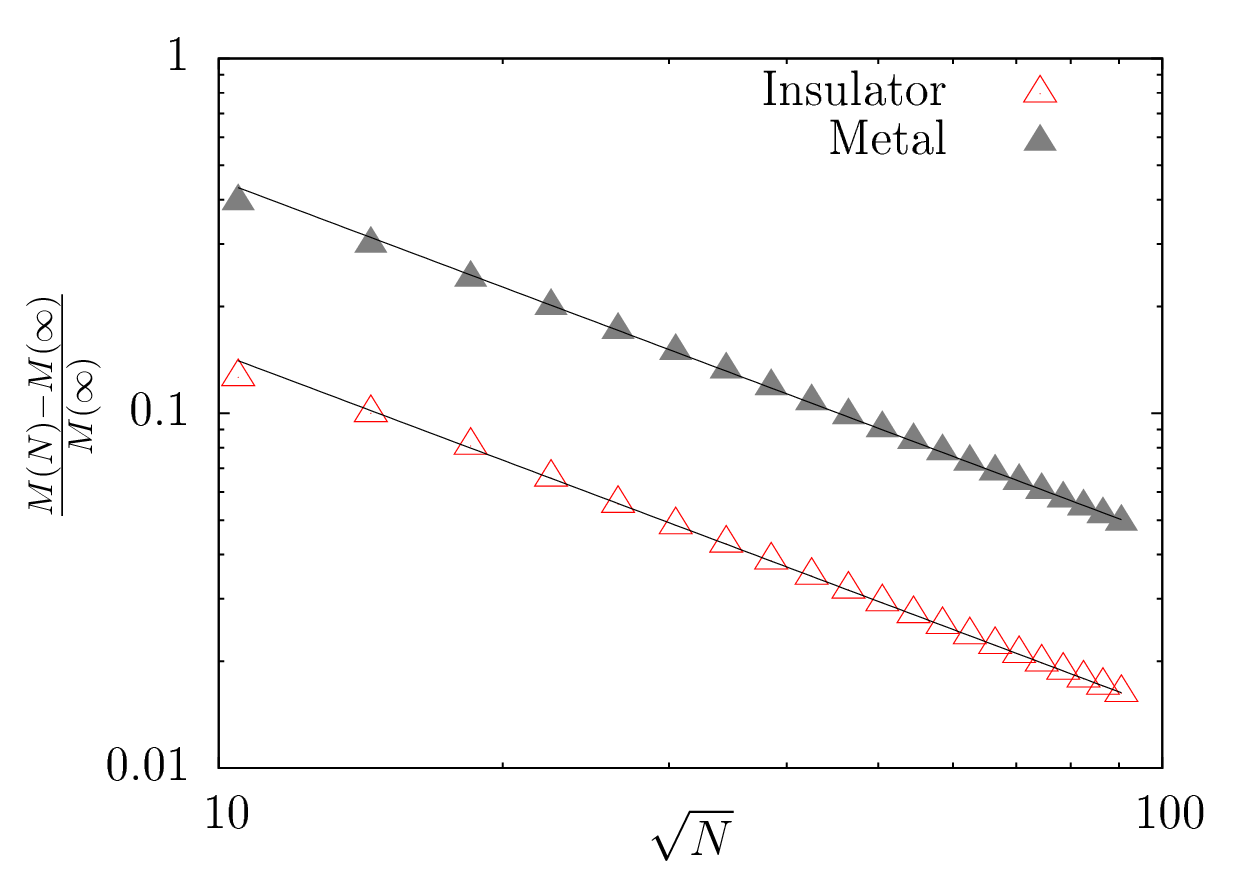}
\caption{(color online). Convergence with flake size of the standard formula, \eqs{def}{def1}, in log-log scale; a typical metallic ($\mu = -1.7$ in the valence band) and a typical insulating ($\mu = -0.7$ at midgap) case are shown. The interpolating straight lines clearly show the $1/L$ convergence.
}
\label{fig:sweep2} \end{figure}

For orientation, we start showing in Fig. \ref{fig:sweep1} the converged magnetization $\M$ of our ``Haldanium'' flake as a function of $\mu$ over the whole range: $\M$ depends on $\mu$ in the metallic range and stays constant while $\mu$ sweeps the gap \cite{chern}. Next, in our metallic test case we set $\mu = -1.7$, rather far from the band edges (see Fig. \ref{fig:sweep1}): we therefore have a sizeable Fermi surface (Fermi loop in 2D), which in turn guarantees a nonzero Drude weight. As recognized by Haldane himself, this model system is a good paradigm for the anomalous Hall effect in metals \cite{Haldane04}. Our simulations also confirm that the OBCs localization tensor diverges with the flake size \cite{rap_a31,Antimo}.

We show next the convergence of the textbook definition in Fig. \ref{fig:sweep2}. We switch to an obvious vector notation and we evaluate 
\[ \M(N)= \frac{\m}{A} = \frac{1}{A} \int_{\rm flake} d \r \; \MM(\r) , \label{textbook}
\] for $N$-site flakes: this is clearly identical to \eqs{def}{def1}. The log-log plot shows that $[M(N) - M]/M$ is proportional to $1/\sqrt{N}$, i.e. to the inverse linear dimension $L^{-1}$ of the flake. Notably, this occurs for both insulating and metallic flakes.

\begin{figure}[t]
\centering
\includegraphics[width=.8\columnwidth]{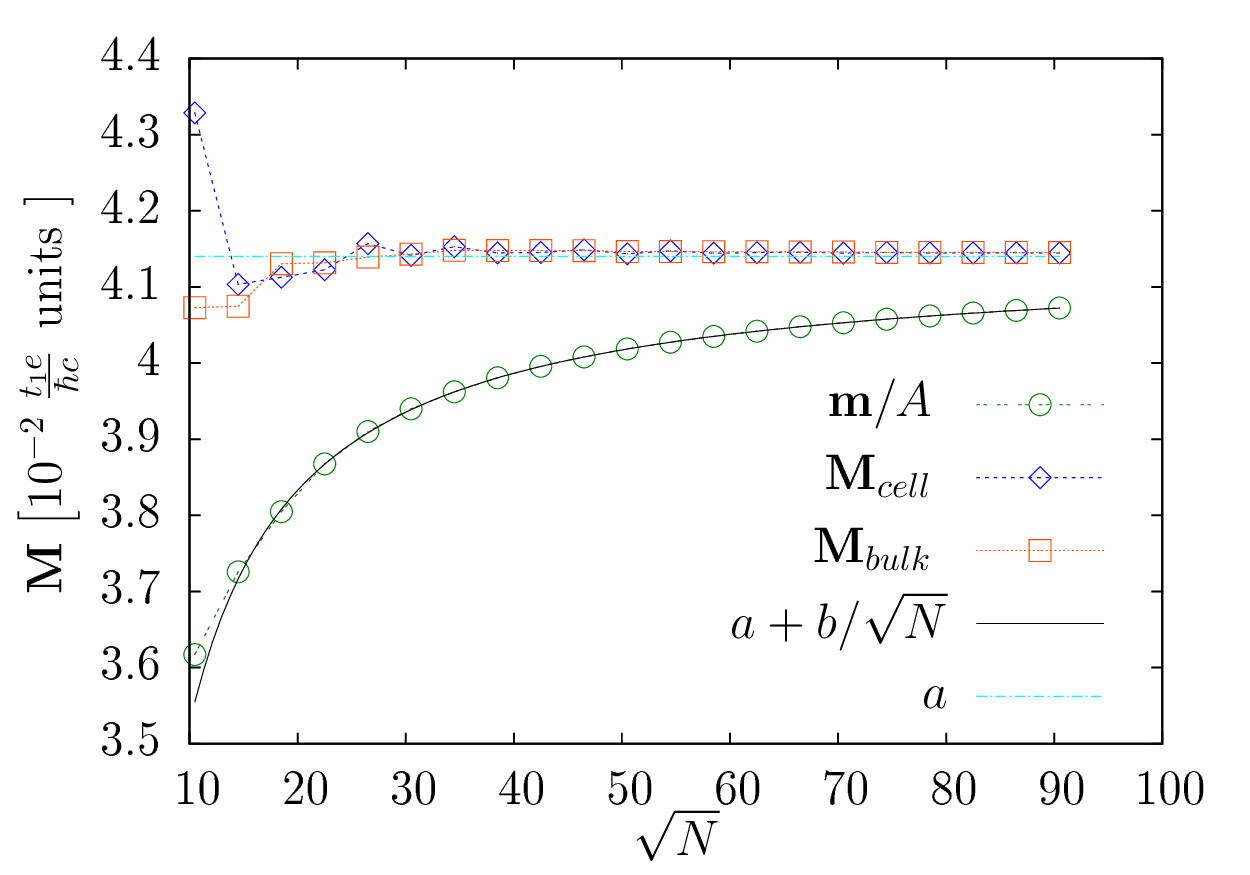}
\caption{(color online). Magnetization as a function of the flake size, at constant aspect ratio in the metallic case: $\mu=-1.7$ in the valence band.}
\label{fig:met} \end{figure}

Our main aim is to assess the locality of $\M$. We therefore compare $\M(N)$, \equ{textbook}, to our {\it local} expressions \[ \Mc = \frac{1}{\ac} \int_{\rm cell} \!\!\!\! d \r \; \MM(\r) , \quad \Mb = \frac{1}{\ab} \int_{\rm bulk} \!\!\!\! d \r \; \MM(\r), \label{local} \] where $\MM(\r)$ is integrated either on a single cell in the center of the flake, or on an inner rectangular region of area 1/4 of the total (see Fig. \ref{fig:flake}). Within our tight-binding Hamiltonian, \equ{local} amounts to averaging either over two sites or over $N/4$ sites.
The results for a typical insulating and metallic case are shown in Figs \ref{fig:ins} and \ref{fig:met}: they show once more that $\m/A$, \equ{textbook}, converges to the bulk $\M$ value as $L^{-1}$. Instead, computations of either $\Mb$ or $\Mc$ by means of our local formulas converge to the bulk value much faster. Remarkably, this happens in {\it both} the insulating and metallic cases. This provides evidence our major claim, i.e. that even in metals the macroscopic magnetization $\M$ can be expressed in terms of the one-body density matrix in the bulk of the sample, disregarding what happens at its boundary.

\begin{figure}[t]
\centering
\includegraphics[width=.8\columnwidth]{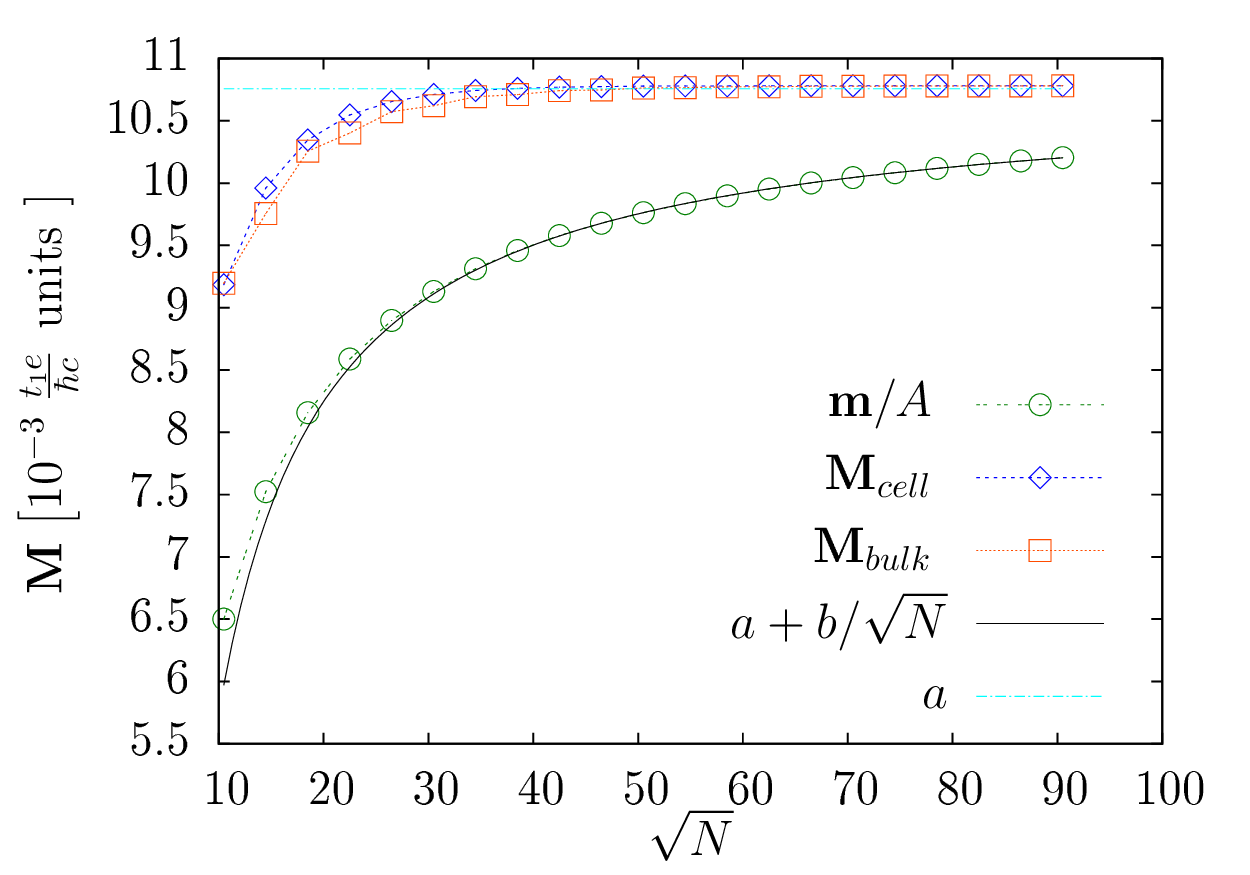}
\caption{(color online). Magnetization as a function of the flake size, at constant aspect ratio in the insulating case: $\mu=-0.7$ at midgap.}
\label{fig:ins} \end{figure}

\begin{figure}[b]
\centering
\includegraphics[width=.8\columnwidth]{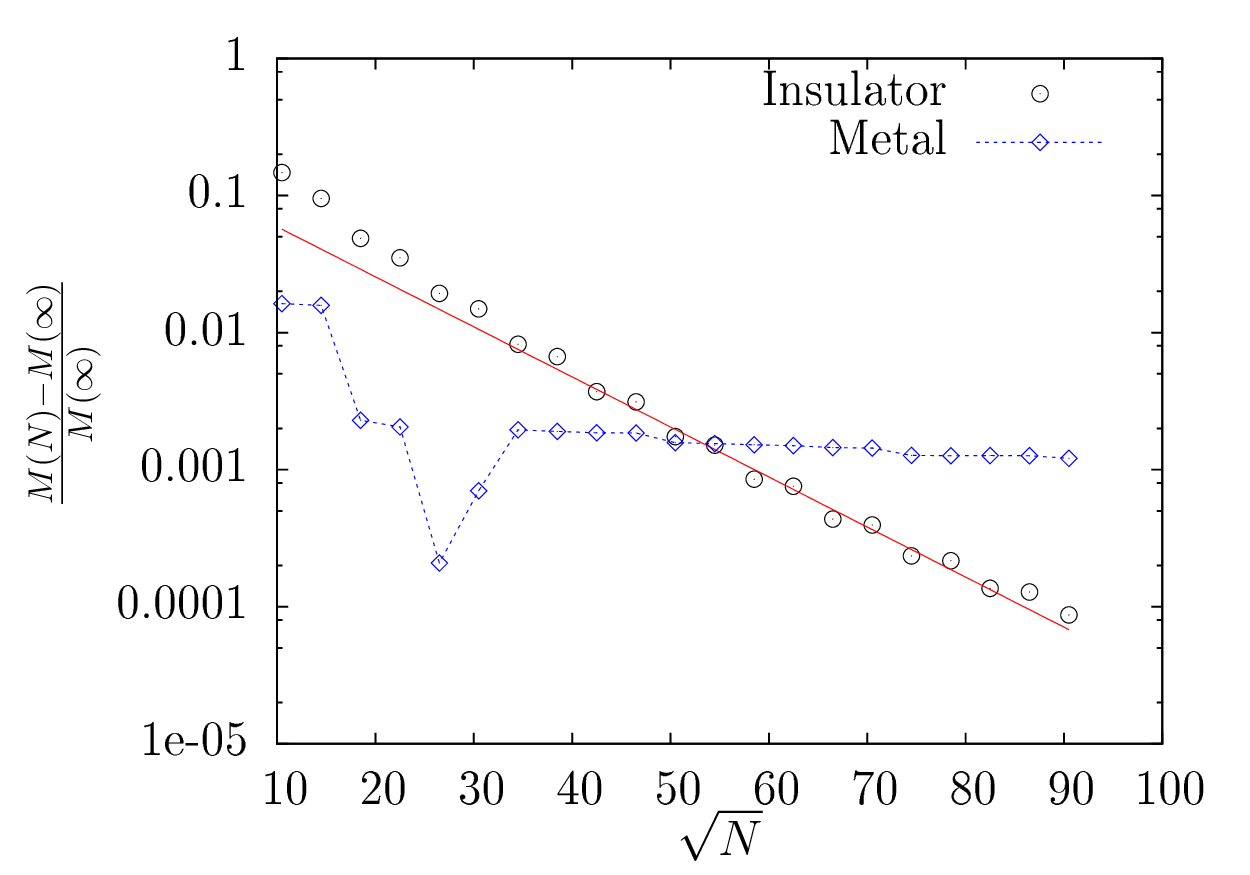}
\caption{(color online). Convergence of magnetization as a function of the flake size (same $\Mb$ as in Figs. \ref{fig:ins} and  \ref{fig:met}) in a log scale. The interpolating line shows an exponential convergence of $\Mb$ in the insulating case, while the convergence is slower in the metallic case.
}
\label{fig:log} \end{figure}

Nonetheless, we also expect the convergence to be {\it qualitatively} different in the two cases: in order to magnify this, we plot both (insulator and metal) on a log scale in Fig. \ref{fig:log}. The plots show that \equ{local} does indeed converge exponentially to the bulk $\M$ value in the insulating case. In the metallic case, instead, the convergence is
definitely slower than exponential. It is not easy to assess the kind of convergence in the metallic case. We may only claim---based on several results such as those shown in Figs. \ref{fig:met} and \ref{fig:log}---that the convergence is of the order $L^{-\alpha}$, with $\alpha$ definitely larger than 1.

Next, we switch to magnetization in a finite macroscopic $\B$ field. Here our main requirement, namely that we are dealing with a 2D metal, is much more delicate. Even if we choose a system that is a very good metal at $\B=0$, the ubiquitous presence of Landau levels (LL) opens gaps in the density of states (DOS) and the metallic nature of our model system must be carefully checked. We therefore rely on some previous results from the literature, where the metallic nature of the model Hamiltonian has been checked by independent means. Following Ref. \cite{Sheng97}, we adopt a simple square lattice with nearest-neighbor interaction, setting $t=1$ in the following; a $\B$ flux $\phi$ equal to $\phi_0/8$---where $\phi_0 = e/(hc)$ is the flux quantum---is included via Peierls substitution. 

\begin{figure}[t]
\centering
\begin{minipage}[b]{.50\linewidth}
\includegraphics[width=\linewidth]{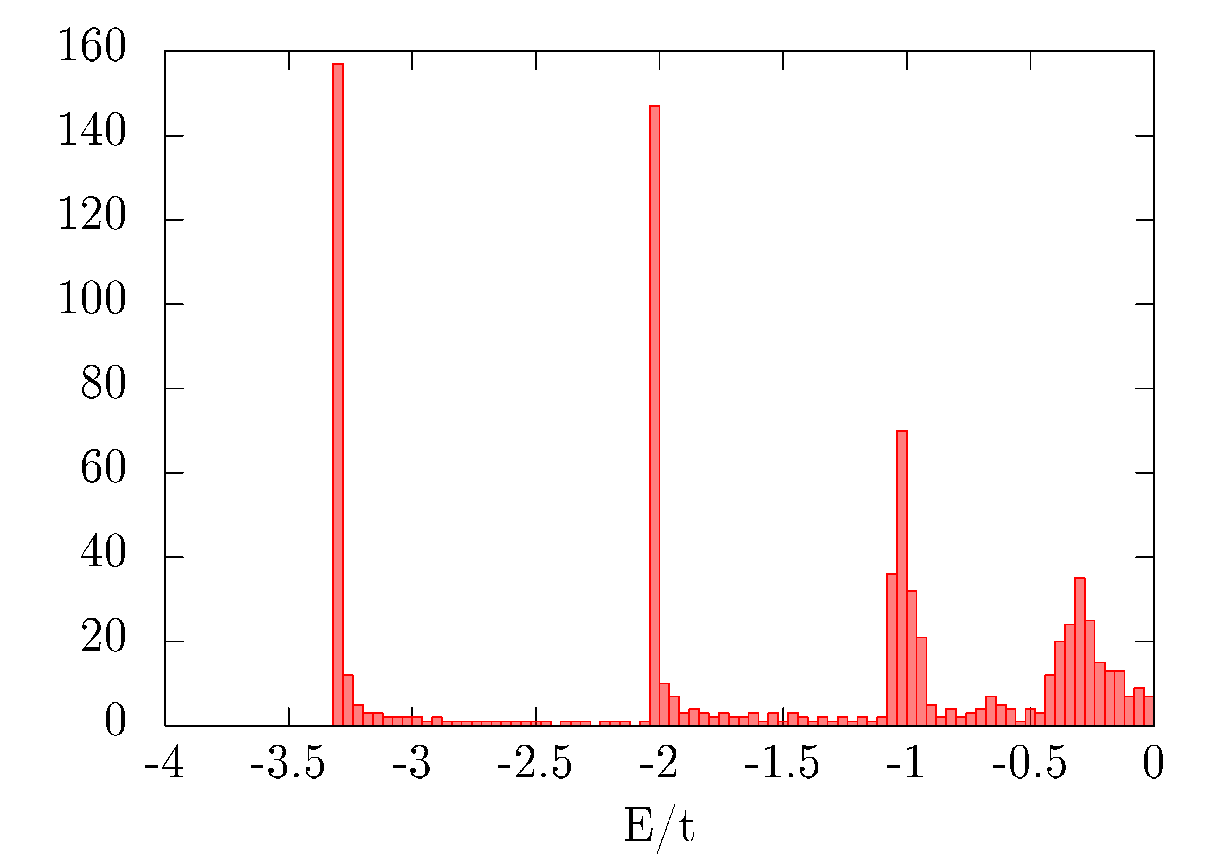}
\end{minipage}\hfill
\begin{minipage}[b]{.50\linewidth}
\includegraphics[width=\linewidth]{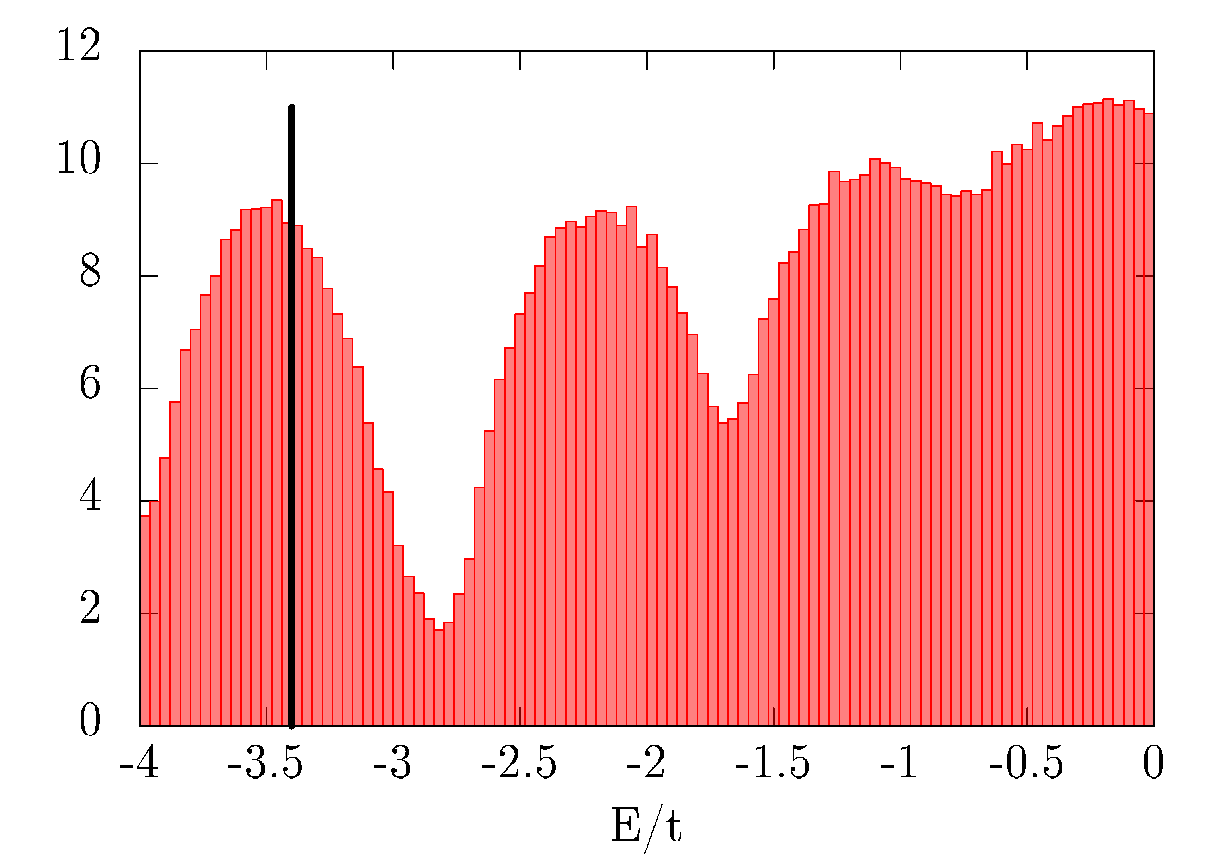}
\end{minipage}
\caption{(color online). Density-of-states hystogram for a 1600-site square flake in a $\B$ field, lower half of the band. Left: pristine (crystalline) flake. Right: disordered flake (see text), average over 100 random configurations; the $\mu = -3.4$ value adopted in the simulations is also indicated. The $\B$ flux per site is $\phi_0/8$, where $\phi_0 = e/(hc)$ is the flux quantum.
}
\label{fig:dos} \end{figure}

The DOS is shown in Fig. \ref{fig:dos}. In the pristine sample (left panel) the LL broadenings are due to both the periodic potential and the finite size of our flake. Ideally, the system is metallic only when the Fermi level $\mu$ is set precisely within a level, and is always insulating otherwise: a condition difficult to fulfill, either experimentally or computationally. We therefore add disorder in order to broaden the metallic regions. Notice that we are indeed mimicking what happens in realistic quantum Hall samples: by increasing the disorder, the regions of extended states around each LL first broaden, then shrink and eventually disappear. Following Ref. \cite{Sheng97} we have added a random onsite term $w_i$ with $|w_i| \leq  W/2$. Setting $\mu = -3.4$ the states at the Fermi level do not localize up to $W \simeq 4 - 5$; we perform our simulation at $W=3$, where our sample is metallic.
Our disorder broadened DOS is shown in Fig. \ref{fig:dos}, right panel.

\begin{figure}[t]
\centering
\includegraphics[width=.8\columnwidth]{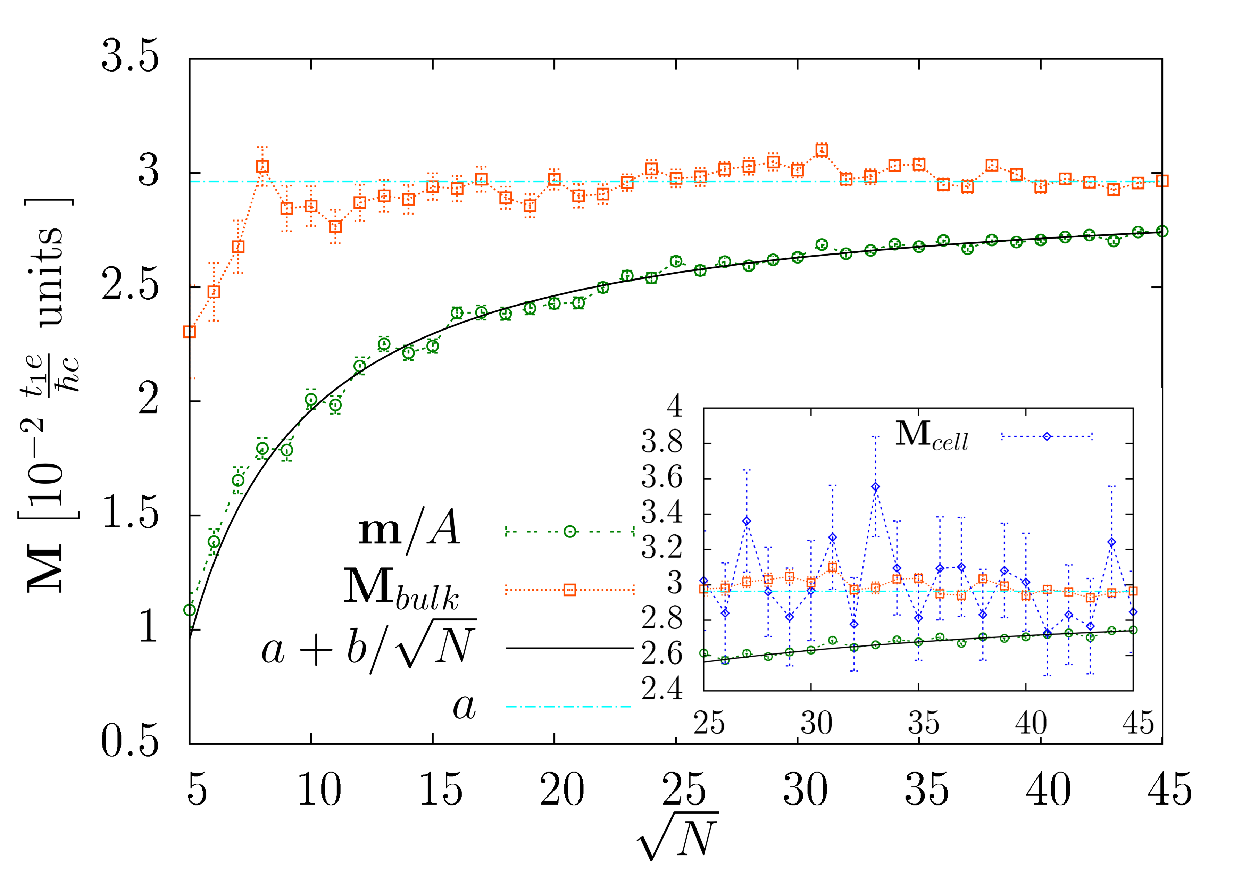}
\caption{(color online). Magnetization as a function of the flake size, for square metallic 2D disordered samples in a macroscopic $\B$ field (see text and Fig. \ref{fig:dos}). The plots show the average over 100 random configurations. $\M_{\rm bulk}$ is the average over the $N/4$ central sites, while $\M_{\rm cell}$ (shown in the inset) is the central-site $\MM$ value.}
\label{fig:mag_a} \end{figure}

Our main results in a $\B$ field are shown in Fig. \ref{fig:mag_a}: as expected, all plots converge to the $\M$ macroscopic value. Comparing the two local formulas, \equ{local}, we notice that $\Mc$ still has large oscillations at our maximum computed size, while such oscillations are quenched by averaging over a larger region, as in $\Mb$. The plot perspicuously shows that $\Mb$ converges to the macroscopic $\M$ value much faster than the textbook definition, \eqs{def}{def1}. The test case dealt with here (square lattice at $\B \neq 0$) behaves similarly in this respect to the Haldane model at $\B=0$, presented above.

In conclusion, the present simulations provide evidence that the $\M$ value in a uniformly magnetized metal---in either zero or nonzero macroscopic $\B$ field---can be retrieved by accessing the bulk electron distribution (i.e. the one-body density matrix) in the bulk region of the sample only. Tinkering with the boundary does not alter the $\M$ value: this is a virtue of our approach which is {\it not} based on currents. The transformation from \equ{def} to \equ{exo} shares the same virtue of an integration by parts: the contribution of the boundary currents in \equ{def} is reshuffled into the bulk in \equ{exo}, where any current has disappeared.
Electron localization is qualitatively different in insulators and in metals (exponential vs. power law); despite this important difference, $\M$ behaves qualitatively in the same way in both cases: the electron distribution in the boundary region does not affect the $\M$ value. Our simulations were performed---for the sake of simplicity---over paradigmatic 2D models only; we nonetheless expect that the main message from our simulations carry over with no qualitative change to realistic 3D metallic systems. We also stress that our local approach to orbital magnetization (in either insulators or metals) allows addressing even noncrystalline and/or macroscopically inhomogenous systems (i.e. heterostructures).

\bigskip

A.M. acknowledges a scholarship from Elettra-Sincrotrone Trieste S.C.p.A. and Collegio Universitario per le Scienze ``Luciano Fonda'';
R.R. acknowledges support by the ONR Grant No. N00014-12-1-1041.

\vfill\vfill


\begin{thebibliography}{10}
\bibitem{chern}
{ With our parameter choice, the Chern number is zero}.
\end{thebibliography}

\end{document}